\documentclass[pop,numberedappendix,twocolappendix,iop]{aeb_emulateapj_2015}

\usepackage{amsmath}
\usepackage{color}
\usepackage{longtable}
\usepackage{ulem,array}
\usepackage{grffile}

\usepackage{hyperref}
\hypersetup{
  bookmarks=false,
  pdfpagelabels=false,
  hyperfootnotes=false,
  hyperindex=false,
  pageanchor=false,
  colorlinks=true,
  linkcolor=blue,
  citecolor=blue,
}

\usepackage{multirow}
\usepackage{amssymb}

\makeatletter
\newcommand{\xRightarrow}[2][]{\ext@arrow 0359\Rightarrowfill@{#1}{#2}}
\makeatother

\usepackage[mathscr]{euscript}

\newcommand{\mathbfit}[1]{\textbf{\textit{#1}}}

\renewcommand{\vec}[1]{\mathbfit{#1}}
\newcommand{\bs}[1]{\boldsymbol{#1}}
\newcommand{\rmn}{\mathrm}

\newcommand{\M}{\mathcal{M}}
\newcommand{\T}{\tilde{T}}

\newcommand{\alf}{Alfv$\acute{\text{e}}$n} 

\DeclareSymbolFont{matha}{OML}{txmi}{m}{it}
\DeclareMathSymbol{\varv}{\mathord}{matha}{29}

\SetSymbolFont{symbols}{bold}{OMS}{cmsy}{b}{n}
\DeclareSymbolFont{bmisymbols}{OML}{cmm}{b}{it}
\DeclareMathSymbol{\bvarv}{0}{bmisymbols}{"1D}

\usepackage{array}
\newcolumntype{M}[1]{>{\centering\arraybackslash}m{#1}}
\newcolumntype{N}{@{}m{0pt}@{}}

\begin{document}

\author{Mohamad Shalaby\altaffilmark{1,*} }
\author{Rouven Lemmerz\altaffilmark{1}}
\author{Timon Thomas\altaffilmark{1}}
\author{Christoph Pfrommer\altaffilmark{1} }

\altaffiltext{}{Leibniz-Institut f{\"u}r Astrophysik Potsdam (AIP), An der Sternwarte 16, 14482 Potsdam, Germany}

\email{*mshalaby@live.ca}

\title{The mechanism of efficient electron acceleration at parallel non-relativistic shocks}
\shorttitle{Electron acceleration at parallel non-relativistic shocks}
\shortauthors{Shalaby et al.}

\begin{abstract}

  Thermal electrons cannot directly participate in the process of diffusive acceleration at electron-ion shocks because their Larmor radii are smaller than the shock transition width: this is the well-known electron injection problem of diffusive shock acceleration.
Instead, an efficient pre-acceleration process must exist that scatters electrons off of electromagnetic fluctuations on scales much shorter than the ion gyro radius.
The recently found intermediate-scale instability provides a natural way to produce such fluctuations in parallel shocks.
The instability drives comoving (with the upstream plasma) ion-cyclotron waves at the shock front and only operates when the drift speed is smaller than half of the electron {\alf} speed.
Here, we perform particle-in-cell simulations with the SHARP code to study the impact of this instability on electron acceleration at parallel non-relativistic, electron-ion shocks.
To this end, we compare a shock simulation in which the intermediate-scale instability is expected to grow to simulations where it is suppressed. In particular, the simulation with an {\alf}ic Mach number large enough to quench the intermediate instability shows a great reduction (by two orders of magnitude) of the electron acceleration efficiency.
Moreover, the simulation with a reduced ion-to-electron mass ratio (where the intermediate instability is also suppressed) not only artificially precludes electron acceleration but also results in erroneous electron and ion heating in the downstream and shock transition regions.
This finding opens up a promising route for a plasma physical understanding of diffusive shock acceleration of electrons, which necessarily requires realistic mass ratios in simulations of collisionless electron-ion shocks.

\end{abstract}


\section{Introduction}
\label{sec::intro}

We observe our Universe primarily though thermal and non-thermal radiation as well as by means of atomic line transitions. The thermal component probes gas (including ionized and neutral components) 
in thermal equilibrium, while the non-thermal emission is emitted by particles such as leptons (e.g. electrons and positrons) 
or hadrons (e.g., protons and ions of heavier elements) that are not in thermal equilibrium.
Thanks to multi-wavelength observations of various astrophysical objects such as supernova
remnants \citep[SNRs,][]{2012Helder}, active galactic nuclei \citep{2010Abdo},
gamma-ray bursts \citep{2009Abdo}, galaxy clusters \citep{Brunetti2014,2019vanWeeren}, and galaxies \citep{Beck2015}, observational insights
into the emission processes and the radiating particle
distributions in collisionless astrophysical plasmas is now possible.
Observed non-thermal emission in our Universe spans a vast range of energies: from radio to TeV gamma-ray energies, and occurs in 
various astrophysical objects spanning many length scales: from sub-AU to Mpc scales.
The observed non-thermal power-law emission indicates that these plasmas are far from thermal equilibrium 
and suggests an efficient particle acceleration process giving rise to a population of cosmic rays
(CRs) composed of electrons, protons, and heavier ions \citep[][]{1987Blandford,Marcowith2016}. In our Galaxy, the acceleration of 
particles near SNR shocks is thought to be the main mechanism for producing CRs up to particle energies of $10^{16}$eV \citep[e.g.,][]{Gaisser2016}.

The interaction of plasmas on both sides of the shock wave produces electromagnetic perturbations.
Protons and heavier ions, with speeds larger than a few times the shock speed, interact resonantly with these electromagnetic perturbations and, as a result, are transported diffusively in the vicinity of shock waves.
Because these magnetic perturbations are converging at the shock front, this leads to proton and ion acceleration through a process known as diffusive shock acceleration \citep[DSA,][]{Krymskii1977,Axford1977,Blandford1978,Bell1978a,Bell1978b}. 
The level of magnetic fluctuations determines the acceleration efficiency of particles, and vice versa, 
accelerated particles excite plasma instabilities \citep{Bell2004} which further amplify the magnetic field near shock fronts.

Observational evidences for electron acceleration at shocks are numerous. These include puzzling giant radio shocks that are observed at the outskirts of merging galaxy clusters.
These diffuse radio-emitting structures are often coincident with weak merger-driven shocks \citep{2019vanWeeren}, and interpreted as synchrotron emitting regions powered by relativistic CR electrons that underwent the DSA process \citep{Ensslin1998,Pfrommer2008,Kang2012,Pinzke2013}.
Moreover, X-ray observations show that the non-thermal emission component at the edges of SNRs such as SN 1006 is due to accelerated CR electrons emitting synchrotron radiation \citep{Willingale+1996}.

The particular morphology of SN 1006 in radio and X-rays has raised the question about the impact of the orientation of the upstream magnetic field on the acceleration mechanism of CR electrons. Radio polarization observations suggest a large-scale magnetic field that is aligned along the north-east to south-west direction \citep{Reynoso2013}. Hence, the magnetic field seems to be aligned with the polar caps visible in radio \citep{Dyer2009}, X-rays \citep{Cassam-Chenaie2008} and TeV gamma rays \citep{Acero2010}, suggesting preferentially quasi-parallel acceleration \citep{Voelk2003}. Azimuthal variations of X-ray cutoff energies \citep{Rothenflug2004, Katsuda2010} and the radius ratio of the forward shock to the tangential discontinuity \citep{Cassam-Chenaie2008} reinforce this picture. Three-dimensional magneto-hydrodynamics (MHD) simulations support the interpretation of preferred quasi-parallel shock acceleration of electrons \citep{Bocchino2011,Schneiter2015,Winner+2020} and protons \citep{Winner+2020,Pais2020a,Pais2020b}.

However, we still lack a detailed understanding of how electrons are accelerated to high energies at these shocks: for electrons and protons traveling with the same non-relativistic speed, the kinetic energy of electrons is much smaller than that of protons owing to their substantially smaller mass in comparison to protons. Hence, unlike protons, these electrons can not directly scatter off of the magnetic perturbations near shock waves and DSA is not possible for electrons without some process(es) that can provide and sustain pre-acceleration of incoming upstream electrons.
This is known as the electron injection problem in shocks \citep{Amano+2010,Guo+2015}.
The importance of understanding the acceleration mechanism of electrons is also reinforced by their much higher radiative yield in comparison to that of protons at the same energy because the Thompson cross section scales inversely with particle mass squared.

In general, electron pre-acceleration proceeds in two steps: 1.\ electrons are heated in the downstream region to energies much higher than their initial kinetic energy, and 2.\ an acceleration phase where electrons scatter off of electromagnetic waves with a wavelength shorter than the shock width, which further increases the electron energy and enables them to participate in the DSA process.
The nature of these waves, which accelerate electrons in step 2, depends on the relative angle, $\theta_{B}$, between the direction of the large-scale background magnetic field and the direction of the propagation of the shock \citep[see, e.g.,][]{Guo+2015}.
Moreover, recent global MHD modeling of the multi-wavelength emission from the supernova remnant SN 1006 shows that efficiency of electron acceleration at quasi-parallel ($0^{\circ}<\theta_{B} < 45^{\circ}$) shocks has to be at least an order of magnitude higher in comparison to quasi-perpendicular ($45^{\circ}<\theta_{B} < 90^{\circ}$) shock configurations \citep{Winner+2020}.

In quasi-perpendicular shocks, electron pre-acceleration is thought to occur via shock drift acceleration (SDA) where electrons are
reflected and accelerated via a combination of the cross shock electric potential and the shock rest frame motional electric field \citep{Wu1984,Krauss+Wu1989}
and/or shock surfing acceleration (SSA) where electron acceleration occurs when they  interact with electrostatic waves at the shock front \citep{Shimada+2000,Xu2020,KumarReville2021}.
Theoretically, the SDA mechanism is shown to be inefficient for accelerating electrons at planar shocks in the so-called scattering-free limit \citep{Ball+Melrose_2001}. However, in the presence of strong pitch-angle scattering, this mechanism is modified to become stochastic SDA, which could result in efficient acceleration \citep{Katou+2019}. 
On the other hand, particle-in-cell (PIC) simulations have shown that the SSA mechanism mediated by waves that are generated due to the Buneman instability is efficient in pre-accelerating electrons \citep[see, e.g.,][]{Bohdan2019}. While these simulations used unrealistically low ion-to-electron mass ratios and (too) high {\alf} speeds, it remains to be shown how these results can carry over to more realistic astrophysical conditions.

For quasi-parallel shocks, it is usually assumed that hot electrons generate whistler waves that lead to pitch-angle scattering and acceleration of these electrons. Efficient wave production in this scenario requires high values of {\alf}ic Mach number, $\M_{\rm A}$. However, when tested with PIC simulations, this scenario did not result in any significant electron acceleration \citep{Riquelme+2011,Niemiec+2012}. 
Recently, \citet{sharp2} discovered a new instability (called the intermediate-scale instability) that drives comoving ion-cyclotron waves unstable in the vicinity of the shock front. This presents {\it a new mechanism for generating waves} that can scatter electrons and potentially enables an efficient DSA mechanism for electrons. Unlike the whistler-mediated case, this mechanism requires low values of $\M_{\rm A}$ which is a condition for the new instability to operate. 

In this paper we test this mechanism using 1D3V PIC simulations (i.e., one spatial and three velocity-space dimensions) of parallel electron-ion shocks and show that it indeed leads to very efficient electron acceleration.
The paper is organized as follows.
In Section~\ref{sec::setup}, we present the setup for our simulations, and compute the linear growth rates for the expected unstable wavemodes at the shock front region.
The growth of these wavemodes is responsible for the formation of the shock and for creating non-thermal populations of electrons and ions.
In Section~\ref{sec:Bamp}, we present the evolution of density and magnetic field amplification  in our simulations.
We study the impact of the intermediate-scale instability driven wavemodes on the acceleration of electrons in the downstream region of the shock in Section~\ref{sec:nth1}.
In Section~\ref{sec:thermal}, we discuss the heating of the shocked plasmas in the downstream and the shock front region. We also compare this heating to analytical heating models at shocks.
The fraction of energy that is channeled into non-thermal  electrons and ions is quantified and its evolution is presented in Section~\ref{sec:nothermal}. We discuss our findings and their implications in Section~\ref{sec:dis}, and conclude the paper in Section~\ref{sec:concl}.
Throughout this paper, we assume the SI system of units.

\section{Non-relativistic shock simulations}
\label{sec::setup}

\begin{figure}
\includegraphics[width=8.8cm]{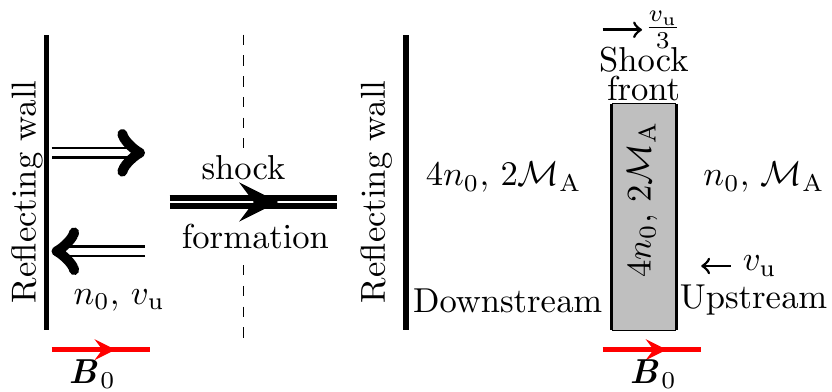}
\caption{The formation of a strong parallel shock in the contact discontinuity rest frame (simulation frame).
Left-hand side: the reflection of an incident electron-ion plasma drifting to the left side of the simulation with speed $\varv_u$ leads to the formation of the shock.
Right-hand side: after shock formation, the shock speed relative to the upstream plasma is $\varv_{\rm sh} \sim 4 \varv_u/3$ (assuming a strong shock and adopting the Rankine-Hugoniot jump conditions), and the density jump at the shock front is on average $\sim 4 n_0$, where $n_0$ is the number density of the far upstream plasma.
That is, at the shock front the {\alf} speed of ions $\varv_{\rm A}$ is $0.5$ of that in the far upstream, which means that the {\alf}ic Mach number, $\M_{\rm A}$, in this region is twice of that in the far upstream.
\label{fig:ShockSketch}}
\end{figure}

Here we discuss the setup for our shock simulations and compute the scales and the growth rates for the expected unstable linear wavemodes at the shock front regions.

\subsection{Simulation setup}

We perform 1D3V particle-in-cell (PIC) simulations using the SHARP code \citep{sharp,resolution-paper,sharp2}, where the shock is formed by reflecting the upstream electron-ion plasma (initially moving towards the left, i.e., negative $\bs{\hat{\vec{x}}}$ direction) at $x=0$.
That is, the shock is formed by the interaction of two exact replica of the same plasma moving in opposite directions, and the simulations are performed in the rest-frame of the contact discontinuity. A sketch for the initial configuration and the resulting shock formation is shown in Figure~\ref{fig:ShockSketch}.

In all simulations, the upstream plasma skin-depth is resolved by 10 computation cells, and the CFL stability condition on the computational time step, $\Delta t$, is such that $c \Delta t = 0.45   ~ \Delta x$, where $c$ is the speed of light in vacuum.
We use 200 particles per species in each computational cell.
The boundary of the computational domain expands with time on the right-hand side to save computational cost while containing all energetic particles in the precursor of the shock.
The upstream  plasma is initially moving with velocity $\varv_u = -0.1 c$, and both electrons (with mass $m_e$) and ions (with mass $m_i$) have the same physical temperature  $T_e = T_p = 4 \times 10^{-8} m_i c^2/k_{\rm B}$, where $k_{\rm B}$ is Boltzmann's constant.
Initially, both species have the same uniform number density $n_i=n_e=n_0$.

For such shocks, the expected shock speed (in the rest-frame of upstream plasma) is 
$\varv_{\rm sh} = 4 |\varv_u| /3 \sim 0.133 c$.
Therefore, the sonic Mach number $ \M_s = \varv_{\rm sh} / \varv_s \sim 
365$, where the sonic speed is $\varv_s = \sqrt{\Gamma_{\rm ad} k_{\rm B} (T_e+T_p)/m_i 
}$, and  $\Gamma_{\rm ad} = 5/3$ is the adiabatic index. That is, all shock simulations presented here have high sonic Mach number.
The initial (large-scale) background magnetic field $\vec{ B}_0 = B_0  \bs{\hat{\vec{x}}}$ is such that the {\alf}ic Mach number $\M_{\rm A} = \varv_{\rm sh}/\varv_{\rm A}$, where $\varv_{\rm A} = B_0/\sqrt{ \mu_0 n_i m_i}$ is the {\alf} speed of ions.

It is important to note here that collisionless shocks found in the intracluster medium have $\M_s \sim 1$ to 3 \citep{Ryu+2003, Pfrommer+2006,Vazza+2009,Schaal+2016}, and for such low values of the sonic Mach number, the intermediate-scale growth rates are much larger in comparison to that at the gyro-scale \citep{sharp2}. This implies a stronger impact of the intermediate-scale unstable modes on the acceleration of electrons in the intracluster medium.
However, we leave a demonstration of this point in simulations to future works.

\subsection{Unstable linear wavemodes at the shock front}
\label{sec::theory}

\begin{figure}
\includegraphics[width=8.6cm]{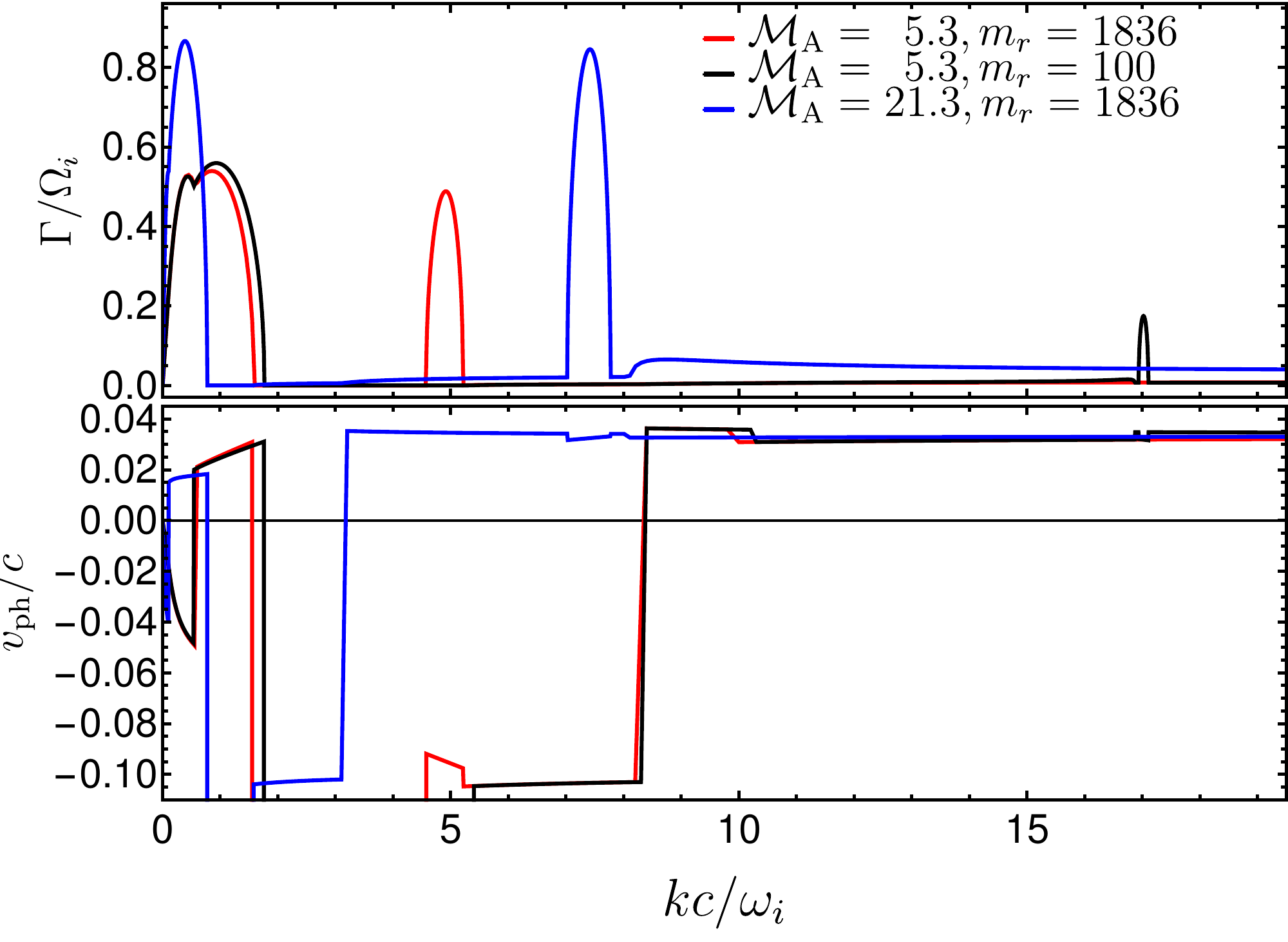}
\caption{Top: growth rates ($\Gamma$) normalized to the non-relativistic ion gyro-frequency, $\Omega_i$.
Bottom: the phase velocity ($\varv_{\rm ph}$) of the fastest growing parallel (unstable) electromagnetic wavemodes created by the penetration of upstream cold plasma drifting  into the denser plasma at the shock front (in the simulation frame).
The solutions are shown for $\M_{\rm A} = 21.3$, with a realistic ion-to-electron mass ratio, $m_r=m_i/m_e$ (blue curves).
In the case of $\M_{\rm A} = 5.3$, solutions are shown for a realistic value of $m_r$ (red curves) and a reduced $m_r$ (black curves).
Only the case with $\M_{\rm A}=5.3$ and a realistic mass ratio is expected to excite the intermediate-scale instability~\citep{sharp2} and thus the small scale ion-cyclotron wave modes that are comoving with the upstream plasma with velocity $\varv_u = -0.1 c$ are amplified, i.e., come from solutions of $D^{+}=0$ (see Equation~\ref{eq:disp}).
Small scale unstable wavemodes in the other (blue and black) cases are whistler waves, i.e., they derive from solutions of $D^{-}=0$ (see Equation~\ref{eq:disp}).
\label{fig:GrowthRate}}
\end{figure}

To study the impact of the intermediate-scale instability on the heating and the acceleration of electrons at parallel non-relativistic shocks, we present simulations that differ in their ability to excite the intermediate-scale instability, which manifests itself by exciting ion-cyclotron waves that are comoving with the upstream plasma~\citep{sharp2}.
The upstream drifting plasma drives wavemodes unstable at the shock front\footnote{In the linear dispersion calculation we present here, the background number density is assumed to be uniform.
Number density non-uniformity could change the growth rates if the non-uniformity scale is close to the wavelengths  of the unstable modes.
However studying this theoretically, even in the linear regime, is tedious \citep[see, e.g.,][]{sim_inho_18, th_inho_20}.}. 
Assuming gyrotropic momentum distributions for various species, the 
dispersion relation, in the simulation frame, for parallel (w.r.t.\ the background magnetic field $B_0$) propagating electromagnetic wavemodes with wavenumber $k$ and complex frequency 
$\omega$ can be written as~\citep{Schlickeiser+2002}
\begin{eqnarray}
D^{\pm}
&=& 1- \frac{k^2c^2}{\omega ^2} 
\nonumber \\
&&+
\sum_{s=1}^{4}
\frac{ \omega_s^2 }{ \gamma_s  \omega ^2 }
\left[
\frac{ \omega -k \varv_{\parallel,s} }
{  k \varv_{\parallel,s}  - \omega  \pm \Omega_s}
-
\frac{ \varv_{\perp,s}^2  \left(k^2c^2-\omega ^2\right) /c^2 }
{2 \left(k \varv_{\parallel,s}  -\omega \pm \Omega_s \right)^2}
\right] ~\hspace{-0.16cm} .
\label{eq:disp}
\end{eqnarray} 
Here, $s=\{1,2\}$ denote the shocked electron and ion plasmas at the shock front, respectively, and $s=\{3,4\}$ denote the incoming (upstream) cold electron and ion plasmas, respectively. The drift speeds are $\varv_{\parallel,s} = \{ - \varv_{u}/3, - \varv_{u}/3, \varv_{u}, \varv_{u} \}$, 
and the relativistic gyro-frequencies $\Omega_s = \{- m_r \Omega_0, 
\Omega_0, -m_r \Omega_0, \Omega_0\}/\gamma_s$, where 
$\Omega_0 = e B_0/m_i $ is the non-relativistic ion gyro-frequency, and 
the Lorentz factor is $\gamma_s = (1 - \varv_{\parallel,s}^2 - \varv_{\perp,s}^2)^{-1/2}$.

Solutions to $D^{+}=0$ ($D^{-}=0$) are solutions for the left (right) handed polarized electromagnetic wavemodes.  When solving for the maximum growth rates at some $k$, we solve the dispersion relation with both signs, and find the fastest growing mode regardless of its polarization.
For various species $s$, the perpendicular speeds are $\varv_{\perp,s} = \varv_s/\sqrt{2} ~ \forall s $,
the plasma frequencies are $\omega_s = \{ \sqrt{3 m_r} ~\omega_i, 
\sqrt{3}~ \omega_i, \sqrt{m_r} ~\omega_i, \omega_i \}$, where $\omega^2_i= e^2 n_i/(m_i \epsilon_0)$ is the square of the ion plasma frequency in the far upstream of the shock, $e$ is the elementary charge, and $\epsilon_0$ is the permittivity of free space. 

We present a simulation that can excite the intermediate-scale instability at the shock front region. It has an upstream {\alf}ic Mach number $\M_{\rm A} = 5.3$ and uses a realistic mass ratio $m_r = m_i/m_e=1836.$
That is, the shock front {\alf}ic Mach number 
\begin{eqnarray}
\M^f_{\rm A} \sim 10.6 < \sqrt{m_r}/2 \sim  21.4 .
\end{eqnarray}
This condition represents the necessary criteria for driving such comoving ion-cyclotron wave modes \citep{sharp2}.
The solution of the dispersion relation (Equation~\ref{eq:disp}) for 
this case is shown in Figure~\ref{fig:GrowthRate} (red curves).
The lower panel shows that the phase velocity of the unstable wave modes (at $kc/\omega_i \sim 5$) are comoving with the upstream drifting plasmas ($\varv_{\rm ph} \sim -0.1 c$) for the simulation with $\M_{\rm A} = 5.3$ and $m_r=1836$ (red curve).
To demonstrate the importance of this instability in electron-heating, we present two more simulations where the condition of this instability is violated.

\begin{deluxetable}{ cccccc }
\tablewidth{8.7cm}
\tabletypesize{\footnotesize}
\tablecolumns{5} 
\tablecaption{
Parameters of our electron-ion parallel shock simulations \vspace{-0.2cm}
\label{table:Sims}
}
\tablehead{ Name & $\varv_u/c$\tablenotemark{a} &  $\M_{\rm A}$\tablenotemark{b}
&   $\M_s$\tablenotemark{c}
&    $m_i/m_e$ 
& Condition\tablenotemark{d} 
}
\startdata
\rule{0pt}{12pt}  Ma5Mr1836  & -0.1 & 5.3 & 365 &  1836 & \checkmark 
\\
\rule{0pt}{12pt}  Ma5Mr100	 & -0.1 & 5.3 & 365 &  100	  & $\times$ 
\\
\rule{0pt}{12pt}  Ma21Mr1836 & -0.1 & 21.3 & 365 & 1836 & $\times$
\enddata
\tablenotetext{a}{Upstream plasma velocity in the contact-discontinuity rest frame.}
\tablenotetext{b}{Upstream {\alf}ic Mach number.}
\tablenotetext{c}{Upstream sonic Mach number.}
\tablenotetext{d}{This shows whether the condition ($\varv_{\rm sh}/\varv_{\rm A} < \sqrt{m_r}/2 $) for exciting the intermediate-scale instability at the shock front is satisfied.}
\end{deluxetable}

The first simulation has $\M_{\rm A} = 21.3$, with a realistic mass ratio, $m_r=1836$.
In this case, $\M_{\rm A}^f \sim 40.6 > \sqrt{m_r}/2 \sim  21.4$. 
Indeed, the solutions of the dispersion relation for this case (blue curves in Figure~\ref{fig:GrowthRate}) show that sub ion-skin-depth unstable whistler wavemodes are  not comoving with the upstream plasma and thus can be easily quenched.
For this simulation, we increase $\M_{\rm A}$ by decreasing $B_0$, i.e., by lowering $\varv_{\rm A}$.

The second simulation has $\M_{\rm A} = 5.3$, but with a reduced mass ratio, $m_r=100$. In this case, $\M_{\rm A}^f \sim 10.3 > \sqrt{m_r}/2 =  5 $.
Again, the solutions of the dispersion relation for this case (black curves in Figure~\ref{fig:GrowthRate}) show that the sub ion-skin-depth unstable whistler wavemodes are not comoving with the upstream plasma and thus can also be easily quenched.
For this simulation, we increase $T_e=T_i$ such that $T_i/m_i$ and thus the sonic speed $\varv_s$ is unchanged. 
Consequently, the sonic Mach number $\M_s$ is also unchanged.

A summary of parameters of the three simulation that we present here is given in Table~\ref{table:Sims}.
It is important to note that larger values of the number density at the shock front, i.e., $n > 4n_0$, increases $\M_{\rm A}^f$ in simulations. Hence, we have added a safety margin to our chosen values of $\M_{\rm A}^f$ that can in principle account for on additional factor of four enhancement in density over its MHD prediction without changing our conclusion on the condition of whether the intermediate-scale instability is excited in our simulations, as given in Table~\ref{table:Sims}.

\section{Shock formation and magnetic field amplification; self-driven scattering centers}
\label{sec:Bamp}

\begin{figure*}
\includegraphics[width=18.5cm]{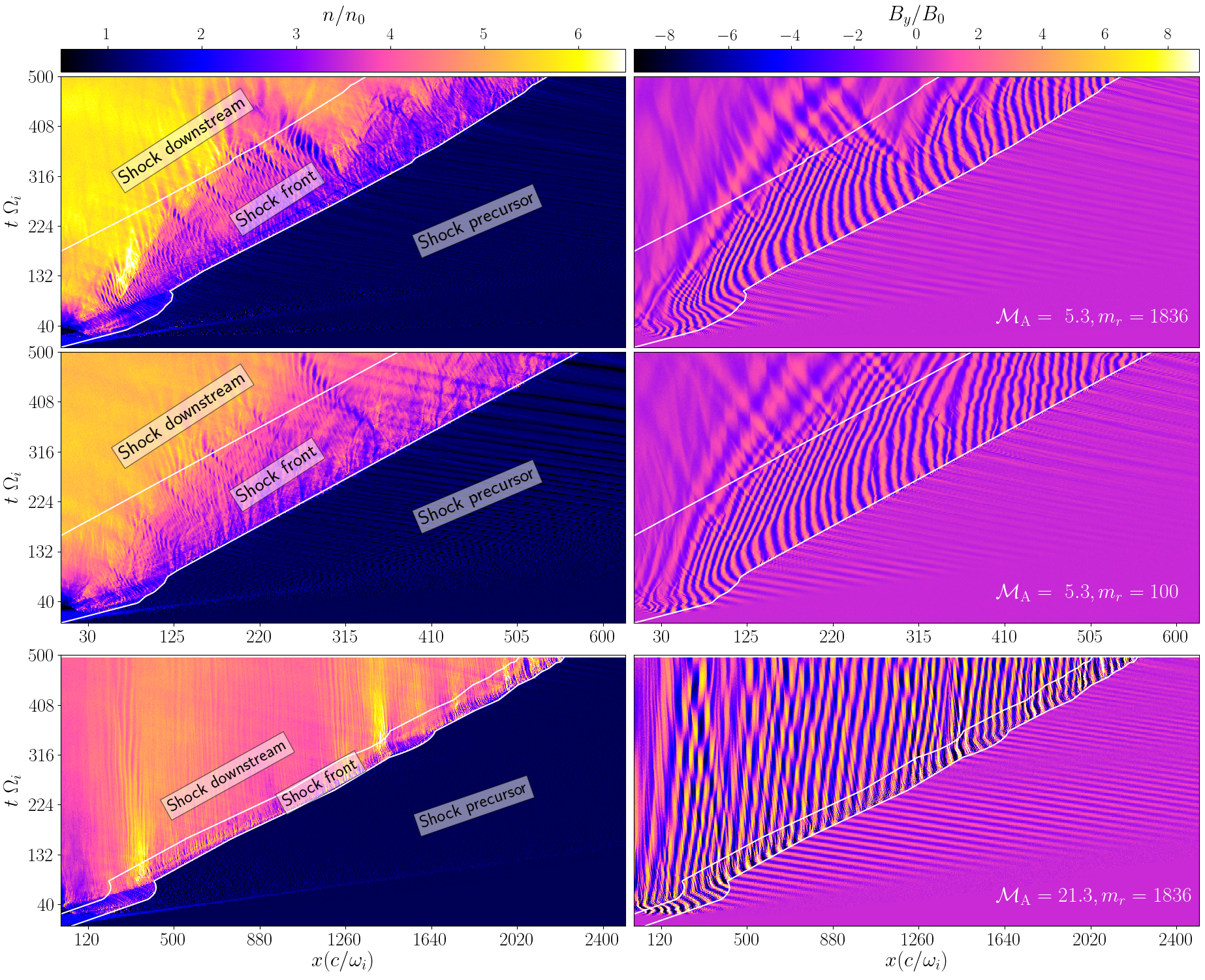}
\caption{Left: evolution of the number density normalized to the far upstream number density ($n_0$).
Right: evolution of $B_y$ normalized to the far upstream parallel magnetic field ($B_0$). 
These are the evolution in the simulation Ma5Mr1836 (top),  Ma5Mr100 (middle), and Ma20Mr1836 (bottom). 
\label{fig:XT-evol}
}
\end{figure*}

The interaction of the drifting and reflected plasma results in instabilities on scales both longer and shorter than the ion-skin depth.
Such unstable modes slow down the reflected plasma and thus create an over-density ($n> 2 n_0$) behind the shock transition region\footnote{Here, $n$ is the number density of both ions and electrons together.}, where $n_0$ is the far upstream number density.
After the formation of the shock, the interaction of the incoming upstream plasma with the denser plasma behind the shock also forms an unstable plasma configuration.
This leads to particle scattering and a constant heating and acceleration of the upstream plasma as it is swept over by the shock.
The left-hand side of Figure~\ref{fig:XT-evol} shows the time evolution of the number density, $n$, normalized to $n_0$, in all the simulations.
The right-hand side shows the time evolution of $B_y$ normalized to the initial background (parallel) magnetic fields $B_0$ (the $B_z$ evolution is very similar to that of $B_y$, albeit shifted slightly in space).
Figure~\ref{fig:XT-evol} shows the formation of the shock via the excitation of unstable magnetic field wavemodes in various simulations.
After the formation of the shock, wavemodes on small and large scales are unstable as seen in Figure~\ref{fig:GrowthRate}. These modes are continuously driven at the shock front region as the shock swipes through the upstream plasma.

In Figure~\ref{fig:XT-evol}, the white lines that separate the shock front and the upstream regions indicates the location of the shock in various simulations. While the exact position of the white line is determined visually, we note that its slope is approximately given by 
\begin{eqnarray} 
\frac{ \Delta t \Omega_i }{  \Delta x \omega_i/c }  =
\frac{ \Omega_i/\omega_i }{ \varv_{\rm sh}  /c }   
= \frac{ 1}{\M_{\rm A} }.
\end{eqnarray}
That is, the slope for the Ma20Mr1836 simulation is smaller by a factor of $\sim 4$ in comparison to the other simulations, and thus the range in  $x$-direction is larger by the same factor.
We define the shock front (transition) to be a region that is 200 $c/\omega_i$ wide behind the location of the shock, i.e., it is the region between the two white lines in various panels of Figure~\ref{fig:XT-evol}.

As particles (especially ions) scatter back and forth across the shock front (transition) region, they are accelerated to higher energies, and thus some particles escape from this region toward the upstream plasma. 
These excites unstable wavemodes, which  in turn scatter most of these counter-streaming energetic particles back toward the shock front region \citep{Bell1978a}.
This process generates the so called shock precursor region ahead of the shock in the upstream plasma as seen in the right-hand panels of Figure~\ref{fig:XT-evol}.
In the next sections, we closely look at the nature of these driven modes and their impact on particle acceleration and heating for both ions and electrons.

The overall density jump between the far upstream and far downstream region ($n/n_0$) is notably larger than expected from the MHD jump condition that predicts $n/n_0= 4$ (for a strong parallel shock). Instead, our simulations with $\M_{\rm A}=5.3$ show an overall density jump of $n/n_0 \sim 6$.
This has already been seen in a number of PIC simulation before as summarized by \citet{Bret2020}.
In hybrid-PIC simulations, where the jump in the density is not constrained by using a very stiff electron equation of state, 
\citet{Colby+2019} shows that for a simulation with $\M_{\rm A}=20$ the density jump grows to $n/n_0  \gtrsim 5.5$.

\section{Impact of intermediate-scale instability driven wavemodes on acceleration}
\label{sec:nth1}

\begin{figure}
\includegraphics[width=8.6cm]{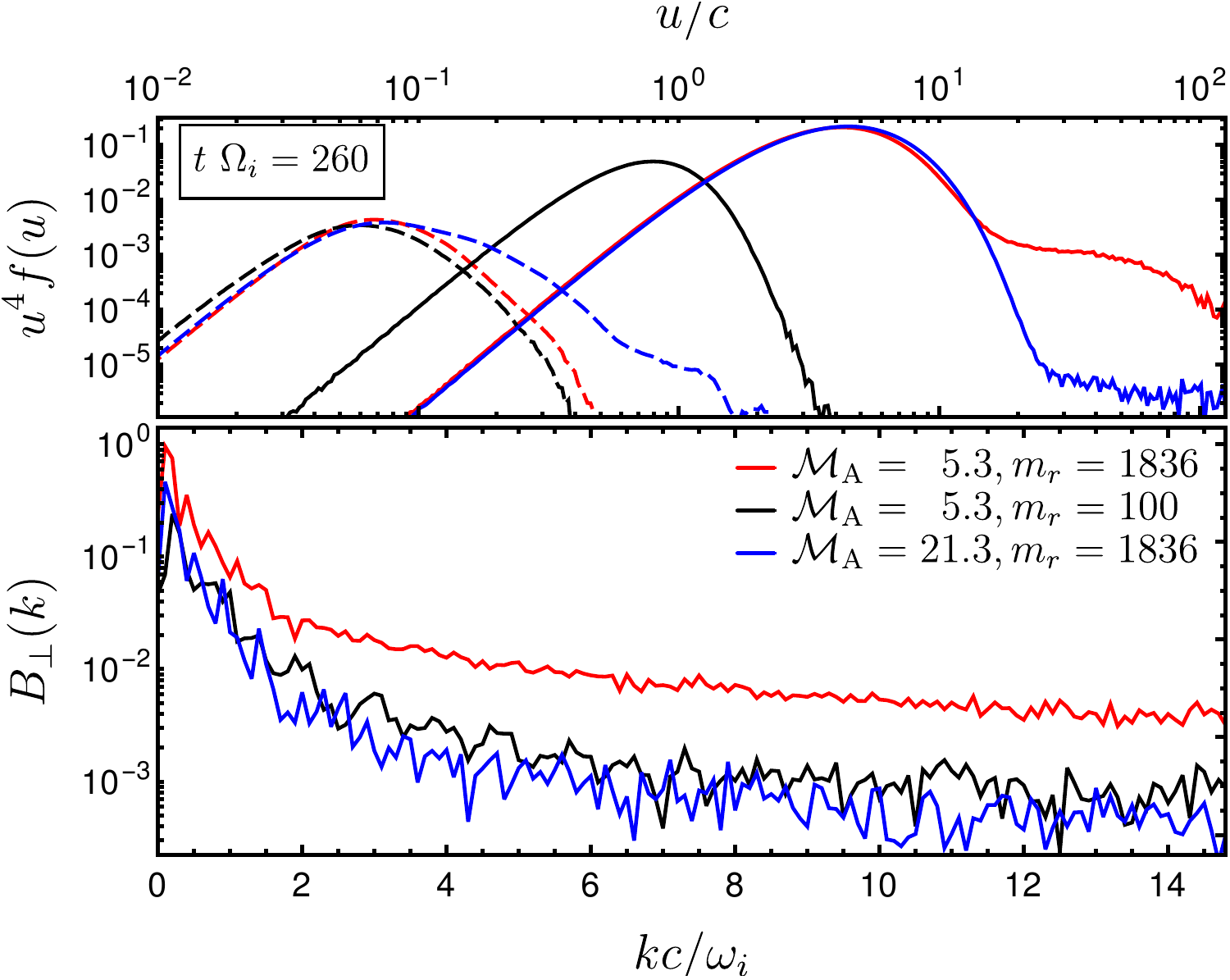}
\caption{Top: downstream electron (solid) and ion (dashed) momentum spectra at $t ~\Omega_i= 260$ from Ma5Mr1836 (red),  Ma5Mr100 (black),  Ma20Mr1836 (blue) simulations.
The downstream is defined to be at a distance that is larger than 200 $c/\omega_i$ from the shock front; see Figure~\ref{fig:XT-evol}.
Bottom: downstream perpendicular magnetic energy in Fourier space at $t ~\Omega_i= 260$ .
This shows that the co-moving (traveling towards the downstream) unstable waves that are driven at the shock front in the Ma5Mr1836 simulation generate a much higher level of small-scale magnetic fields. These increased magnetic fluctuations imply a much stronger scattering and thus, more efficient acceleration of electrons in comparison to the other simulations.
\label{fig:spectra300}}
\end{figure}

\begin{figure*}
\centering
\includegraphics[width=18.5cm]{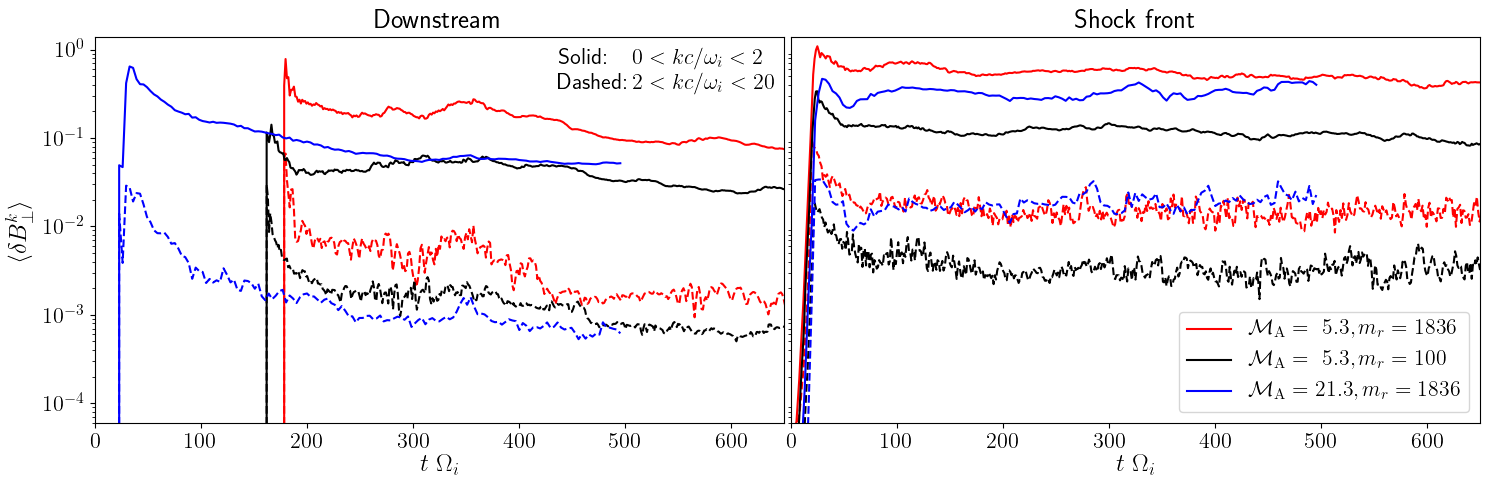}
\caption{Time evolution of the average perpendicular magnetic field power, $\langle \delta B^k_{\perp} \rangle$, in the downstream (left) and shock front (right) regions.
The dashed lines indicate the small-scale ($k c/\omega_i > 2$) time evolution while solid lines show the large-scale ($k c/\omega_i< 2$) time evolution. 
Different simulations are indicated with different colors; Ma5Mr1836 with red,  Ma5Mr100 with black, and   Ma20Mr1836 with blue.
\label{fig:KLS}
}
\end{figure*}

The importance of the intermediate-scale instability on the particle acceleration process can be quantified by comparing simulations which differ in their ability to excite comoving ion-cyclotron modes via the intermediate-scale instability.
As seen in Figure~\ref{fig:GrowthRate}, all simulations are predicted to have unstable wavemodes on scales smaller than the ion skin depth at the shock front (transition) region.
However, only in the simulation Ma5Mr1836, these  modes are comoving with the incoming flow (red, see the bottom panel of Figure~\ref{fig:GrowthRate}), and hence their mode power can be transferred to the downstream region where they can scatter electrons.

The top panel of Figure~\ref{fig:spectra300} shows the electron and proton spectra as a function of dimensionless velocity $u_s/c$. To obtain these spectra, we first transform to the frame in which the plasma is at rest before we compute the momentum spectra from which we estimate the plasma temperature as laid out in Appendix~\ref{app:temp}. 
As shown in the top panel of Figure~\ref{fig:spectra300}, the simulation Ma5Mr1836 (where the intermediate-scale instability can grow) has a much higher efficiency in converting the incoming flow kinetic energy into non-thermal energy in electrons in the downstream region. In fact, it is more efficient (by about 2 orders of magnitude) in accelerating electrons in comparison to the simulation with a higher $\M_{\rm A}$ (blue), suggesting that there is a much more efficient process at work that enables electrons to be efficiently accelerated in our low-$\M_{\rm A}$ model (red).

Moreover, the Ma5Mr100 simulation (black) has the same $\M_{\rm A}$ and $\M_{\rm s}$ as the simulation shown in red but uses a lower mass ratio: $m_r=100$.
As shown in the top panel of Figure~\ref{fig:spectra300}, this results in a much smaller electron acceleration efficiency in this simulation.
Additionally, the heating of electrons in the downstream region of the simulation with a reduced mass ratio (black) is much smaller compared to the simulations with a realistic mass ratio (red and blue)\footnote{In Appendix~\ref{app:temp}, we show that the normalized temperature, $k_{\rm B} T_s/m_s c^2$, of the thermal part of the plasma population of species $s$ is proportional to the value of $u/c$ where the $u^4 f(u)$ is maximized, see Equation~\eqref{Eq:TR}.}.
That is, the top panel of Figure~\ref{fig:spectra300} demonstrates that using a lower mass ratio results not only in an artificially suppressed electron acceleration efficiency but also leads to an erroneous electron heating in the downstream region.

As we will now argue, the differences in the acceleration efficiency of electrons as well as for particle heating at shocks is a direct consequence of the differences in the nature of the unstable small-scale modes.
The bottom panel of Figure~\ref{fig:spectra300} shows that the small-scale power of the perpendicular magnetic field in the downstream region is much larger in the simulation model Ma5Mr1836 in comparison to the simulations Ma5Mr100 (black) and Ma21Mr1836 (blue).
In the latter two cases, the small-scale unstable modes are not comoving with the incoming flow, and thus can be easily quenched, which leads to a much smaller amount of small-scale perpendicular magnetic power in the downstream regions.

To demonstrate the growth of these small scale wavemodes, in Figure~\ref{fig:KLS} we plot the time evolution of the average perpendicular magnetic field for small (dashed) and large (solid) scale wave modes in various simulations. To this end, we take the power of the perpendicular magnetic field as shown in the bottom panel of Figure~\ref{fig:spectra300} and compute the average for larger scales, i.e., for $kc/\omega_i<2$ and for small scales, i.e., for $2<kc/\omega_i<20$.
The top panel of Figure~\ref{fig:GrowthRate} predicts that in the linear regime, both small and large scale magnetic perturbations grow with similar rates in all simulations at the shock transition region. 
The right-hand panel of Figure~\ref{fig:KLS} demonstrates that this prediction is in an excellent agreement with the growth of magnetic perturbations in various simulations at the shock transition region.

On the other hand, the left-hand panel of Figure~\ref{fig:KLS} demonstrates the crucial importance of the difference in the  nature of the grown small scale wavemodes.
As shown in the bottom panel of Figure~\ref{fig:GrowthRate}, only for the Ma5Mr1836 simulation (red), the small-scale wavemodes are comoving with the incoming flow. This means that these modes are not easily quenched in the shock transition region, which is demonstrated by the ability of maintaining a high level of small-scale fluctuations as a result of continued growth of the instability, which explains the much larger efficiency of electron acceleration in this simulations, see Figure~\ref{fig:spectra300}. 
Moreover, much of this small-scale electro-magnetic power is transferred to the downstream region as can be inferred from the left-hand panel of Figure~\ref{fig:KLS}.
In the case of the simulations without the intermediate-scale instability (Ma5Mr100 and Ma21Mr1836), much of the small-scale power is quenched and thus a much smaller fraction of this is transferred to the downstream region of the shock.

\begin{figure*}
\includegraphics[width=18.5cm]{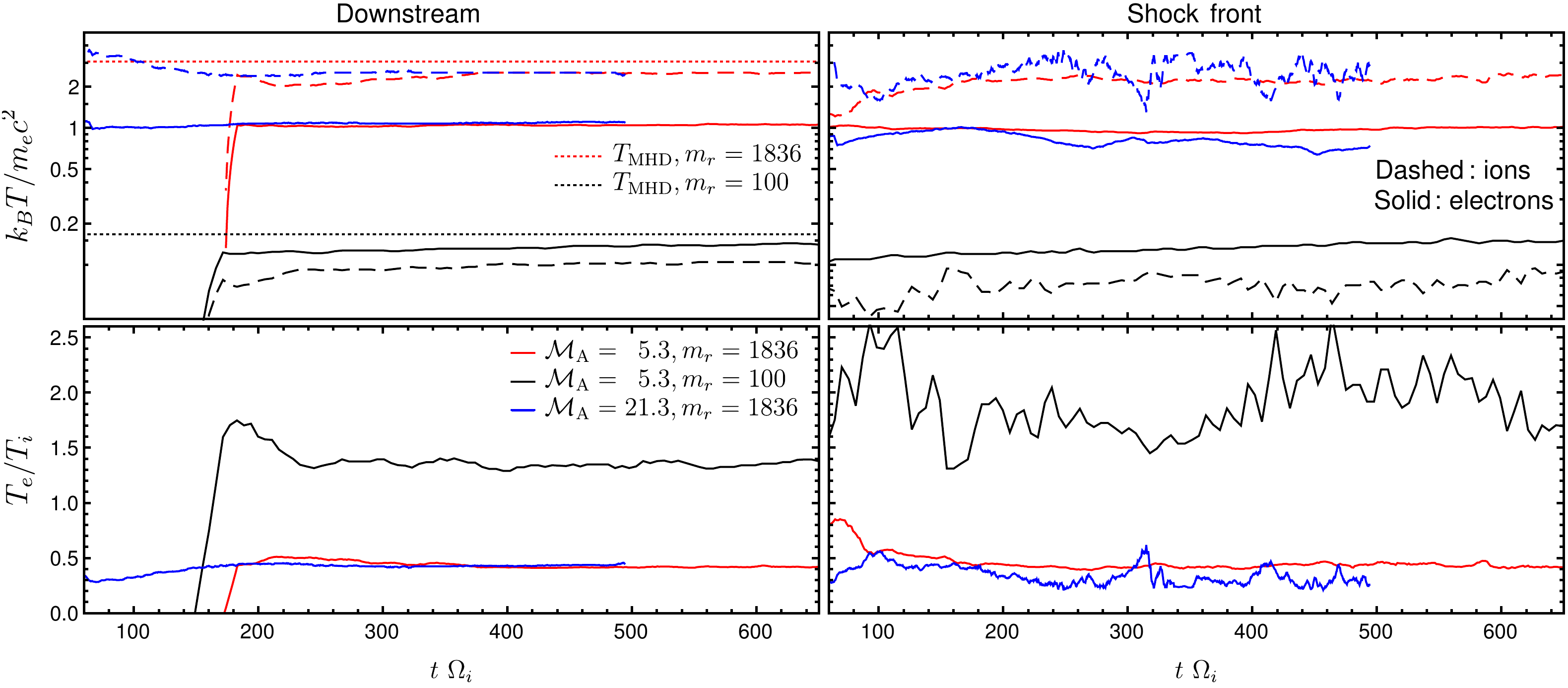}
\caption{Top: shown is the time evolution of downstream (left) and shock front (right) temperatures of ions (dashed) and electrons (solid) in various simulations. Dotted red (black) line in the top left-hand panel indicate the predicted MHD ion temperature for $m_r = 1836 ~ (100)$ as given in Equation~\eqref{eq:kTiV}. 
Bottom: evolution of the electron-to-ion temperature ratio of the plasma in the downstream (left) and shock front (right) regions in various simulations.
\label{fig:Temp}
}
\end{figure*}

In the next sections, we show that the use of a lower mass ratio also leads to erroneous heating and acceleration efficiency  in the shock front (transition) region, and that holds true throughout the evolution of the simulations (as seen in Figures~\ref{fig:Temp} and \ref{fig:Eff}).

\section{Heating of the shocked plasmas}
\label{sec:thermal}

All simulations presented here only have a parallel background magnetic field and are characterized by the same sonic Mach number $\M_s \sim 365$.
After shock formation, electrons and ions in the downstream and shock front regions are heated. Modeling the saturation temperatures for electron and ions  in such collisionless shocks is very important for understanding the observations of SNRs~\citep{Vink+2015}, clusters of galaxies~\citep{Russell+2012}, and the warm hot intergalactic medium which contains a significant fraction of the baryon in the universe~\citep{Bykov+2008}.

\subsection{Models for saturation temperatures}

The MHD model for strong parallel shocks predicts that all the upstream drifting kinetic energy is converted into thermal energy in the post-shock rest frame.
The expected temperature can be computed using the Rankine-Hugoniot jump conditions for a strong, parallel MHD shock in the regime of high sonic Mach numbers, $\M_s\gg1$ \citep[e.g.,][]{boyd}, so that  the temperature of the shocked regimes (downstream and shock front regimes) is given by
\begin{align}
k_{\rm B} T =  \frac{3}{16} \mu m_i \varv_\rmn{sh}^2
  = \frac{1}{3} \mu m_i \varv_u^2 = \frac{1}{600} m_r m_e c^2,
  \label{eq:kTi}
\end{align}
where $\varv_\rmn{sh}=4\varv_u /3$ is the upstream velocity in the shock frame, $\varv_u=0.1c$ is our adopted value for the upstream velocity in the simulation frame, we adopted the adiabatic index $\Gamma_\rmn{ad}=5/3$ appropriate for our case, and $\mu = 1/2$ is the mean molecular weight of an electron-proton fluid.
The MHD assumption implies that the ion and electron temperatures equilibrate on negligible timescales so that $T_e=T_i$ in this model.
That is, we expect the temperature in the shocked regimes to depend on the adopted mass ratio as follows:
\begin{equation}
k_{\rm B}  T_{i} =  k_{\rm B}  T_{e} =  \left\{ 
  \begin{aligned}
    &3.06~~~ m_e c^2
    &&\mbox{  for  } m_r = 1836,\\
    &0.166~ m_e c^2
    &&\mbox{  for  } m_r = 100.\\
  \end{aligned}
  \right.
    \label{eq:kTiV}
\end{equation}
These values are shown with dotted red ($m_r=1836$) and black ($m_r=100$) lines in the top left-hand panel of Figure~\ref{fig:Temp}.

Because these shocks are collisionless, the MHD assumption of strong collisional relaxation is not fulfilled and the electrons may not equilibrate with the ions.
Therefore, another theoretical model for saturation temperatures can be obtained by assuming that thermalization of the initial flow kinetic energy occurs separately for ions and electron \citep{Vink+2015}.
This implies that the electron-to-ion temperature $T_e/T_i = m_e/m_i$, and for our simulations this predicts $k_{\rm B} T_e  = 0.0016 ~ m_e c^2$ (where we used Equation~\ref{eq:kTi}).

\subsection{Saturation temperatures in simulations}

In Figure~\ref{fig:Temp} we show the evolution of the temperature in the downstream (left) and shock front (right) regions in various simulations.
The top panel of Figure~\ref{fig:Temp} shows that a significant fraction of the incoming flow kinetic energy is converted to thermal heating. However, unlike the prediction of MHD shocks, the electron and ion temperatures differ.
In all simulations, the ion temperatures in the downstream and shock front regions saturate slightly below the MHD prediction because a fraction of the incoming flow kinetic energy is channeled into magnetic energy (as seen in Section~\ref{sec:Bamp}) and non-thermal energy of cosmic ray ions and electrons (as we show in Section~\ref{sec:nothermal}).

For ions, we find $k_{\rm B}T_{i}\sim2.5~m_e c^2$ (for $m_r = 1836$, which is only 22.4\% lower than the MHD prediction) and $k_{\rm B}T_{i}\sim0.1~m_e c^2$ (for $m_r = 100$, which is 66\% lower than the MHD prediction).
On the other hand, for electrons, saturation temperatures have a larger mismatch with the MHD predicted temperatures.
 
While the mismatch of the post-shock temperature in the case of a realistic mass ratio can be approximately accounted for by efficient ion acceleration and energy retained in electromagnetic fluctuations, the offset in the case of a mass ratio of $m_r = 100$ is puzzling and cannot be explained by energy in non-thermal components, which only add up to 5\%. We conjecture that the missing plasma instabilities as a result of the artificially small mass ratio precludes exciting sufficient electromagnetic modes that provide the necessary scattering centers for dissipating all the kinetic energy of ions.

Figure~\ref{fig:Temp} also shows that assuming an independent evolution for ions and electrons is not correct since the initial flow kinetic energy (mainly from the ions) excites magnetic field wavemodes which both heat and accelerate electrons to an energy density that, in total,  is much larger than that of the initial flow kinetic energy for the electrons.
The top panel of Figure~\ref{fig:Temp} shows that the saturation temperature for ions (electrons) is lower (larger) than predicted from such models, and the bottom panels show that the prediction for the electron-to-ions temperature ratio of such models is inconsistent with all simulations.

In agreement with the prediction from MHD shocks, simulations with a realistic ion-to-electron mass ratio have saturation temperatures of ions and electrons that are independent of the {\alf}ic Mach number $\M_{\rm A}$. This is demonstrated by the excellent agreement between blue and red (dashed and solid) curves in Figure~\ref{fig:Temp} albeit with $T_e < T_i$.
The saturation temperature of electrons is the same in the downstream and shock front regions, which is also approximately true for ions. This means that all heating occurs in the shock front region and {\it no further} heating occurs in the downstream region for ions or electrons.
The lower panels of Figure~\ref{fig:Temp} show that, at saturation, $T_e/T_i = 0.4$ in the downstream regime for the two simulations that use a realistic mass ratio and $T_e/T_i = 1.4$ for the simulation with $m_r=100$.

In comparison to simulations with a realistic mass ratio, the simulation that uses a low mass ratio of $m_r=100$ results in {\it incorrect} saturation temperatures for ions (by factor of $\sim 25$) and electrons (by factor of $\sim 7.4$) at both, the downstream and shock front regions as shown in the top panel of Figure~\ref{fig:Temp}. 
This low saturation temperature for electrons leads to strong suppression in electron acceleration efficiency as we show in Section~\ref{sec:nothermal}.
Moreover, the use of a lower mass ratio results in a wrong electron-to-ion temperature ratio at saturation, i.e., $T_e>T_i$ as can be seen from the bottom panel of Figure~\ref{fig:Temp}.
Saturation temperatures where $T_e < T_i$ are observed for example in Balmer-dominated shocks~\citep{van-Adelsberg+2008}.
That is, $T_e/T_i>1$ found in the simulation with low ion-to-electron mass ratio disagrees with observed saturation temperature ratios.
Surprisingly, for the simulation with $m_r=100$, the ion and electron temperatures continue to increase from the shock front region to the downstream region. This indicate that further heating occurs in the downstream region of this simulation in disagreement with the results from the simulations with a realistic mass ratio.

\section{Formation of non-thermal particles}
\label{sec:nothermal}

\begin{figure*}
\includegraphics[width=18.5cm]{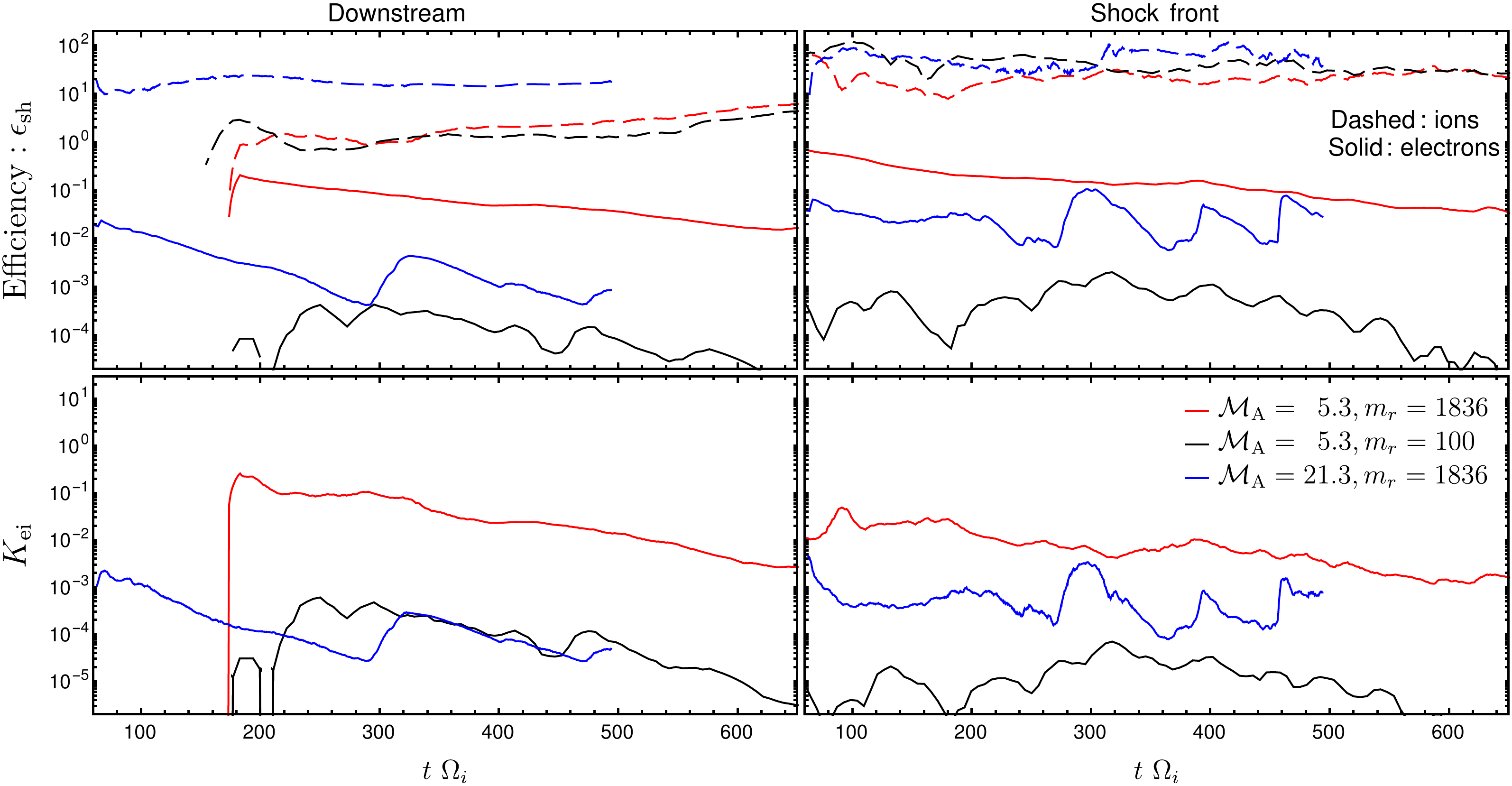}
\caption{Top: the time evolution of the acceleration efficiency (as defined in Equation~\ref{Eq:Eff-sh}) in the downstream (left) and shock front (right) region for ions (dashed) and electrons (solid) in various simulations.
Bottom: the time evolution of the fraction of non-thermal energy of electrons relative to that of ions ($K_{\rm ei}$) in the downstream (left) and shock front (right) regions in various simulations.
\label{fig:Eff}
}
\end{figure*}

As shown above, the upstream kinetic energy is channeled into magnetic energy (Section~\ref{sec:Bamp}), non-thermal energy (Section~\ref{sec:nth1}) and thermal energy (Section~\ref{sec:thermal}) in the downstream and shock front regions of the system. 
Here we quantify the amount of energy carried by non-thermal ions and electrons in these regions in Figure~\ref{fig:Eff}.
In Appendix~\ref{app:temp}, we show that the fraction of particles with non-thermal energy can be quantified in various ways, and here we focus on how much energy density is carried by these particles when compared to the upstream kinetic energy density, i.e., using Equation~\eqref{Eq:Eff-sh}.
We define non-thermal particles to be particles with $u>5u_m$~\citep{Caprioli2014a,Xu2020}, where $u=\gamma \varv$ is the spatial-part of the 4-velocity, $\gamma$ is the Lorenz factor, and $u_m$ is defined as the value of $u$ where $u^4 f(u)$ is maximized.
As an example, particles to the right of the blue vertical lines in Figure~\ref{fig:Thermal} are considered to form the non-thermal part of the momentum distribution. 
This quantification is done for different regions as a function of time and was used to compute the time evolution of the efficiency shown in the top panel of Figure~\ref{fig:Eff}.
The top panel shows the time evolution of the efficiency $\epsilon_{\rm sh}$, i.e., the percentage of upstream kinetic energy density that is channeled into non-thermal electrons and ions in the downstream (right) and shock front regions (left).
By contrast, the bottom panel shows the time evolution of the ratio of non-thermal energy of electrons to that of ions, $K_{\rm ei}$, in various simulations also in the downstream (left) and shock front (right) regions.

The non-thermal energy density of ions (dashed curves) is about 20-30\% of the upstream kinetic energy density, and this result is roughly independent of the ion-to-electron mass ratio $m_r$ at the shock front region.
The simulation with a higher $\M_{\rm A}$ has a slightly higher efficiency in producing non-thermal ions at the shock front (transition) region.
Moreover, in the downstream region, it contains a higher fraction of non-thermal ions indicating that this simulation is more efficient in scattering ions that are accelerated in the shock front (transition) region back to the downstream region.

On the other hand, the fraction that is channeled into non-thermal electrons is strongly dependent on both, $\M_{\rm A}$ and $m_r$.
The top panel of Figure~\ref{fig:Eff} shows that, at the shock front, the simulation with $\M_{\rm A}=5.2$ and a realistic mass ratio efficiently accelerates electrons, leading to 0.05--0.1\% of initial flow kinetic energy converted into non-thermal electrons, while in the simulation with a higher $\M_{\rm A}$ and a realistic mass ratio, a smaller fraction (0.01--0.05\%) of the upstream kinetic energy is channeled into non-thermal electrons.
The simulation with a reduced mass ratio leads to about 2-3 orders of magnitude reduction in electron acceleration efficiency.

Clearly, the simulation where the intermediate-scale instability is allowed to grow at the shock front region (red) is much more efficient in confining the accelerated electrons by scattering them back to the downstream regions, leading to a much larger non-thermal electron energy in the downstream region in comparison to the simulation with a higher $\M_{\rm A}$ as seen in the left-hand panel of Figure~\ref{fig:Eff}.
Moreover, the simulation with a low mass ratio (black) leads to significant reduction in the energy channeled into non-thermal electrons in the downstream region.

As shown in the lower left-hand panel of Figure~\ref{fig:Eff}, in the downstream region, the simulation that allows for growth of comoving ion-cyclotron waves (red) has a 2-3 orders of magnitude larger non-thermal electron-to-ion energy ratio, $K_{\rm ei}$, in comparison to that in the simulations where the condition for the intermediate-scale instability is not satisfied.

\section{Discussion}
\label{sec:dis}

In this section, we discuss the impact of a finite thermal temperature on the growth of the intermediate-scale instability, followed by a comparison of our proposed electron pre-acceleration mechanism in comparison to the mechanism proposed by \citet{2015phrvl.114h5003p}. Finally, we discuss how our results can be used to possibly infer a lower limit on the amplitude of the magnetic field at shocks of supernova remnants.

\subsection{Impact of finite plasma beta}

The dispersion relation used in Section~\ref{sec::theory} to predict the unstable modes at the shock transition region assumes a cold background of electrons and ions.
Thus, it is of great importance to consider the impact of the finite plasma beta $\beta_s$ on the growth of the intermediate-scale instability, where  $\beta_s \equiv 2 \varv_{{\rm th},s}^2/\varv_{{\rm A},s}^2$, the square of the thermal speed is $\varv_{{\rm th},s}^2 \equiv k_B T_s/m_s$ and $s=e,i$ for electrons and ions respectively.

The dispersion relation studied in \citet{sharp2} and presented in Section~\ref{sec::theory} are computed assuming $\beta_i =\beta_e=0$ while the simulation presented by \citet{sharp2} has $\beta_e = \beta_i = 2$. Interestingly, the growth rates of the analytical derivation in the cold-background limit and the simulations with the finite $\beta$ values show an excellent agreement, thus indicating that there is no impact of  $\beta_e$ or $\beta_i$ on the maximum growth rate of the intermediate-scale instability.

\citet{sharp2} argue that the damping due to thermal ion-cyclotron damping (at $k \varv_{\rm th,i} = (1$--$2) \Omega_i$) could at most impact one peak of the instability if parameters are fine-tuned and thus typically it cannot impact instability growth for typical astrophysical plasma conditions. Similarly, it is strainght-forward to show that this is also true for thermal electron-cyclotron damping because the separation between the two peaks of the intermediate-scale instability does not solely depend on the ion-to-electron mass ratio.

For the shock problem studied in the present paper, there are two very different regimes of the beta factor: a low-beta regime in the pre-shock region and a high-beta regime in the shock transition zone. In the pre-shock region, we have $\beta_e = \beta_i = 1.28\times10^{-4}$ ($\M_{\rm A}=5.3$)  and $\sim 2 \times 10^{-3}$ $(\M_{\rm A} = 21.3$). The case of $\beta_e \gtrsim1$ would imply a much faster growth of the intermediate-scale instability in comparison to the growth rate of the gyro-scale instability because of the dependence of the growth rate on $\varv_\perp$ as given in equation (6) of \citet{sharp2}. Thus, we expect that in this case the instability will have an even larger impact on the mechanism of electron acceleration but postpone a detailed study of this to future works.

At the shock transition zone of the simulations with $m_r=1836$, the electron are hot and the temperature is such that $k_{\rm B} T_e \sim m_e c^2$. Therefore, we obtain $\beta_e = 1.74$ for the $\M_{\rm A} = 5.3$ simulation and $\beta_e = 27.88$ for the $\M_{\rm A} = 21.3$ simulation, i.e., in both cases, the electron beta is larger than one. However, only in the simulation with $\M_{\rm A} = 5.3$, the relative drift is such that the instability condition is fulfilled and thus allows for the growth of the intermediate-scale instability. This allows resonant interactions between electrons and unstable modes and results in much larger acceleration efficiency as shown in the simulation.

\subsection{On the electron acceleration mechanism}

The proposed mechanism for pre-accelerating electrons works by exciting intermediate-scale waves at the shock transition region.
These sub-ion skin-depth (ion-cyclotron) waves are comoving with the incoming upstream plasma (see bottom panel of Figure~\ref{fig:GrowthRate}) and hence are coherently transported to the downstream region (see Figure~\ref{fig:KLS}).
In the downstream and shock transition regions, the hot electrons scatter off of these unstable waves  and the waves reflected at the contact discontinuity (see top-left panel of Figure~\ref{fig:XT-evol}). This leads to the high acceleration efficiency shown in the red simulation (with $\M_{\rm A} = 5.3$) as seen in Figures~\ref{fig:spectra300} and \ref{fig:Eff}, possibly in a first-order Fermi type process.

The simulation with $\M_A=21.3$ (depicted in blue) shows excitation of sub-ion skin-depth (Whistler) waves at the shock front region. However, these modes are not transported to the downstream region (see Figure~\ref{fig:KLS}), which thus results in a much lower electron acceleration efficiency.
That is, this simulation shows that for a parallel shock geometry, electron pre-acceleration due to scattering with Whistler waves has a much lower efficiency in comparison to that in the red simulation (see Figure~\ref{fig:Eff}).

We emphasize that our proposed mechanism for electron pre-acceleration does not depend on ion acceleration.
This is clearly manifested by the large fraction of pre-accelerated electrons before we observe any significant ion acceleration in the  simulation with $\M_{\rm A} = 5.3$, as shown in the top-panel (red curves) of Figure~\ref{fig:spectra300}.
On the other hand, the mechanism proposed by~\citet{2015phrvl.114h5003p} for electron pre-acceleration relies on the amplification of Bell modes in the shock pre-cursor driven by the propagation of accelerated ions and thus the combination of SDA and scattering with Bell modes resulted in electron pre-acceleration and injection in DSA cycles.

\subsection{Applications to supernova remnants}
\label{sec:SNR}
While we have clearly demonstrated the importance of the intermediate-scale instability for thermalizing and accelerating electrons in our PIC simulations, we will now turn to discuss the potential relevance of this instability in observations of astrophysical shocks. Perhaps the cleanest astrophysical object is the supernova remnant SN~1006, which enables testing our ideas of the prevailing plasma instabilities that are responsible for electron scattering and acceleration at the quasi-parallel shock conditions we encounter at the polar cap regions of that remnant shock. \citet{Winner+2020} perform three-dimensional MHD simulations of the Sedov Taylor explosion and evolve the electron distribution function while accounting for magnetic obliquity-dependent electron acceleration. To this end, their method finds and characterizes the MHD shocks, injects a pre-defined electron power-law spectrum into the post-shock region \citep{Winner+2019}, and explores the effects of varying the magnetic amplification factor on the surface brightness maps as well as the multi-frequency spectrum of SN~1006.

Matching the radial emission profiles and the gamma-ray spectrum  requires a volume-filling, turbulently amplified magnetic field of $B\approx35~\mu$G in the downstream region of the parallel shock and that the Bell-amplified magnetic field \citep{Bell2004} is starting to get damped in the further post-shock region \citep[see figure~2 of][]{Winner+2020}. The exact value of the Bell amplification factor $f_\rmn{B}$ barely changes the multi-frequency spectrum so that we obtain a post-shock Alfv\'en velocity of $\varv_\rmn{A}=B f_\rmn{B}/\sqrt{\mu_0\mu m_p n}\approx200~f_\rmn{B}~\rmn{km~s}^{-1}$, where $\mu=1.25$ is the mean molecular weight of the warm interstellar medium, $m_p$ is the proton rest mass, and $n=0.12~\rmn{cm}^{-3}$. While the Bell amplified field is maximized in the immediate post-shock regime \citep{Caprioli2014b}, the turbulently amplified magnetic field keeps rising in the post-shock regime \citep{Ji+2016}, so that it is appropriate to set $f_\rmn{B}=1$ while noting that the turbulently amplifying field on its route to saturation is replacing the Bell-amplified field as it is damping further downstream.

Adopting the lab frame shock velocity of SN~1006 at the current epoch, $\varv_s=3000~\rmn{km~s}^{-1}$, we obtain a post-shock Alfv\'enic Mach number of
\begin{align}
  \M_\rmn{A}=\frac{\varv_s}{\varv_\rmn{A}}=15.2<\frac{\sqrt{m_r}}{2},
\end{align}
which obeys the condition for exciting the intermediate-scale instability and thus should enable efficient electron acceleration at the polar cap regions of the parallel shocks of SN~1006 \citep[see figure~4 of][]{Winner+2020}.

In fact, efficient electron acceleration at parallel shocks enables us to put a lower limit on the post-shock magnetic field, 
\begin{align}
  \label{eq:Bmin}
  B&>2 \varv_s \sqrt{\frac{\mu_0\rho}{m_r A}}=B_\rmn{min}\\
  &= 22.7~\left(\frac{\varv_s}{3000~\rmn{km~s}^{-1}}\right)
  \left(\frac{n}{A\, 0.1~\rmn{cm}^{-3}}\right)^{1/2}\,\mu\rmn{G},
\end{align}
where $m_r$ is the proton-to-electron mass ratio and $A$ is the atomic mass number of the element responsible for driving the intermediate~scale instability. Hence, if the plasma is composed of abundant ${}^{56}\rmn{Fe}$ ions, the minimum post-shock magnetic field is lowered to $3~\mu$G for the same shock parameters.
For heavy ions to dominate the growth of the intermediate instability, we require them to be very abundant because the growth rate, $\Gamma$, of the intermediate-scale instability depends on the number density of ions via $\Gamma \propto n_{\rm Fe}^{1/3}$.

\section{Conclusions}

\label{sec:concl}

In collisionless shocks, electrons are heated to energies much larger than the kinetic energy of upstream  electrons impinging on the shock. However, in non-relativistic shocks, their Larmor radii fall short of those of ions so that another acceleration mechanism is needed to boost their energies to the point where they can actively participate in the process of DSA.
Previously suggested mechanisms, which were based on driving whistler waves unstable by hot downstream or cold upstream electrons, require high values of the {\alf}ic Mach number and were shown to not work in PIC simulations \citep{Niemiec+2012}. 
In this paper we consider a new mechanism for electron pre-acceleration in quasi-parallel, non-relativistic electron-ion shocks that is based on driving ion-cyclotron waves unstable by means of ions that are drifting with their upstream velocity through the shock transition zone. The corresponding intermediate-scale instability \citep{sharp2} only works for low values of the {\alf}ic Mach number at the shock front, $\M_{\rm A}^f < \sqrt{m_i/m_e} \approx 21$, which is the condition for the instability to operate.

We present results from three 1D3V PIC simulations for parallel electron-ion shocks with sonic Mach number $\M_{\rm s} \sim 365$:
the first (red) simulation uses a realistic ion-to-electron mass ratio of $m_r=1836$ and provides favorable conditions for exciting the intermediate scale instability.
The second (black) simulation employs identical physical initial conditions but has an artificially lower value for the ion-to-electron mass ratio, $m_r=100$, which violates the instability condition.
In the third (blue) simulation, the condition is also violated by using a higher value of $\M_{\rm A}$ and $m_r=1836$.
Highlight results include:
\begin{itemize}

\item Only the simulation that grows the intermediate scale instability scatters hot electrons in the downstream off of driven ion-cyclotron waves. We demonstrate that this efficiently pre-accelerates electrons to energies that enable them to participate in DSA cycles and yields a non-thermal power-law distribution of electrons (see Figure~\ref{fig:spectra300}).

\item This effective electron acceleration comes about because the excited ion-cyclotron waves at the shock front are comoving with the upstream plasma and hence, survive advection into the downstream. In consequence, the amplitude of perpendicular magnetic field fluctuations, which are resonantly scattering hot electrons, is substantially increased in comparison to the other simulations that preclude instability growth (see Figures~\ref{fig:spectra300} and \ref{fig:KLS}).

\item The simulation with the higher value of $\M_{\rm A}$ shows a reduction, by more than 2 orders of magnitude, in the efficiency of electron acceleration (see Figure~\ref{fig:Eff}). However, the electrons in the downstream are heated to the same temperature as the red simulation (see Figure~\ref{fig:Temp}).
  
\item The simulation with the lower mass ratio ($m_r=100$) results in much lower heating of electrons in the downstream region and thus an even lower electron acceleration efficiency (see Figures~\ref{fig:Eff} and \ref{fig:Temp}). We conclude that accurate PIC simulations of collisionless electron-ion shocks require realistic mass ratios.
  
\end{itemize}

These findings put the magnetic amplification processes at shocks into the limelight because of the strict instability criterion that favors low values of $\M_{\rm A}$ and as such, is able to relate efficient electron acceleration with a lower limit on the downstream magnetic field strength (see Equation~\ref{eq:Bmin}). Most importantly, this paper provides an important cornerstone in our understanding of diffusive shock acceleration of electrons, and potentially enables us to construct an analytical subgrid model for including electron acceleration into modern MHD models.

\section*{Acknowledgements}

We would like to thank Anatoly Spitkovsky for discussions on various aspects of the simulations. We also acknowledge comments by Damiano Caprioli and the anonymous referee.
M.S., R.L., T.T., and C.P. acknowledge support by the European Research Council under ERC-CoG grant CRAGSMAN-646955.
The authors gratefully acknowledge the Gauss Centre for Supercomputing e.V. (www.gauss-centre.eu) for funding this project by providing computing time on the GCS Supercomputer SuperMUC-NG at Leibniz Supercomputing Centre (www.lrz.de). This research was in part supported by the Munich Institute for Astro- and Particle Physics (MIAPP) which is funded by the Deutsche Forschungsgemeinschaft (DFG, German Research Foundation) under Germany´s Excellence Strategy – EXC-2094 – 390783311.

\begin{appendix}

\section{Distribution of relativistic particles}
\label{app:temp}

Here we review the shape of the momentum distribution of charged particles in thermal equilibrium, and the impact of the existence of a non-thermal tail.
We discuss how it can be characterized in such cases, and different ways to quantify the fraction of density and energy in its non-thermal tail.

\subsection{Average thermal energy}

In thermal equilibrium, the rest-frame (R) isotropic momentum distribution of relativistic particles (with elementary mass $m_s$) is given by the Maxwell-J{\"u}ttner distribution~\citep{Wright1975}
\begin{eqnarray}
\label{Eq:distribution}
f_{\rm th}(\vec{u}) = \frac{ n_R e^{-\gamma/\T_R}  }{4 \pi \T_R K_2(1/\T_R)} ,
\end{eqnarray}
where $K_2$ is the modified Bessel function of the second kind, the Lorentz-factor is $\gamma = \sqrt{1+ \vec{u} \cdot \vec{u}}$, and $\T_R$ is the thermodynamical equilibrium temperature in the rest frame\footnote{
It is worth mentioning here that because the phase-space volume is invariant under Lorentz transformation, this distribution takes the same form in any other frame, e.g., in the laboratory frame \citep{Wright1975}.} normalized by $m_s c^2/k_{\rm B}$, all velocities are normalized with the speed of light $c$, and $n_R$ is the rest-frame number density of the particles.
Therefore, the average (thermal + rest-mass) energy per particle, normalized by $m_s c^2$, is

\begin{eqnarray}
\langle \bar{E} \rangle &=&  \int d^3u \gamma \frac{f_{\rm th}(\vec{u})}{n_R} = 4 \pi  \int_0^{\infty}  du  \frac{u^2 \gamma e^{-\gamma/\T_R}  }{4 \pi \T_R K_2(1/\T_R)}
\nonumber \\
&=&
\frac{1}{ \T_R K_2(1/\T_R)} \int_0^{\infty}  du u^2 \gamma e^{-\gamma/\T_R} 
\nonumber \\
&=&
\frac{-1}{ \T_R K_2(1/\T_R)} \frac{d}{d(1/\T_R)} \int_0^{\infty}  du u^2  e^{-\gamma/\T_R}
\nonumber \\
&=&
\frac{-1}{ \T_R K_2(1/\T_R)} \frac{d ( \T_R K_2(1/\T_R) )}{d(1/\T_R)} 
=3 \T_R +\frac{K_1 \left(1/\T_R \right)}{K_2\left(1/\T_R \right)}.
~~~~~~~ 
\end{eqnarray}
That is, the average thermal energy per particle is
\begin{eqnarray}
\label{Eq:Eth}
\langle E_{\rm th} \rangle = \left( 3 \T_R +\frac{K_1 \left(1/\T_R \right)}{K_2\left(1/\T_R \right)} - 1 \right) m_s c^2.
\end{eqnarray}

\subsection{Equilibrium rest-frame temperature}
For a distribution of particle momenta, we can find $\T_R$ (assuming a thermal equilibrium of most low energy particles) as follows:
\begin{eqnarray}
\label{Eq:du4fu}
\frac{d }{du} \left[  u^4 f_{\rm th}  \right]  =  \frac{ n_R  u^3 \left(4 \sqrt{u^2+1} \T_R-u^2\right)  }{4 \pi \T_R^2 K_2(1/\T_R) \sqrt{u^2+1}}  e^{-\frac{\sqrt{u^2+1}}{\T_R}} .
\end{eqnarray}
That is, using the value of $u$ for which $u^4 f(u)$ is maximum ($u_m$), we can determines $\T_R$ using
\begin{eqnarray}
\label{Eq:TR}
\T_R = \frac{  u_m^2 }{ 4 \sqrt{1+u_m^2}}.
\end{eqnarray}

\subsection{Momentum distribution with non-thermal particles}

\begin{figure}
\includegraphics[width=8.6cm]{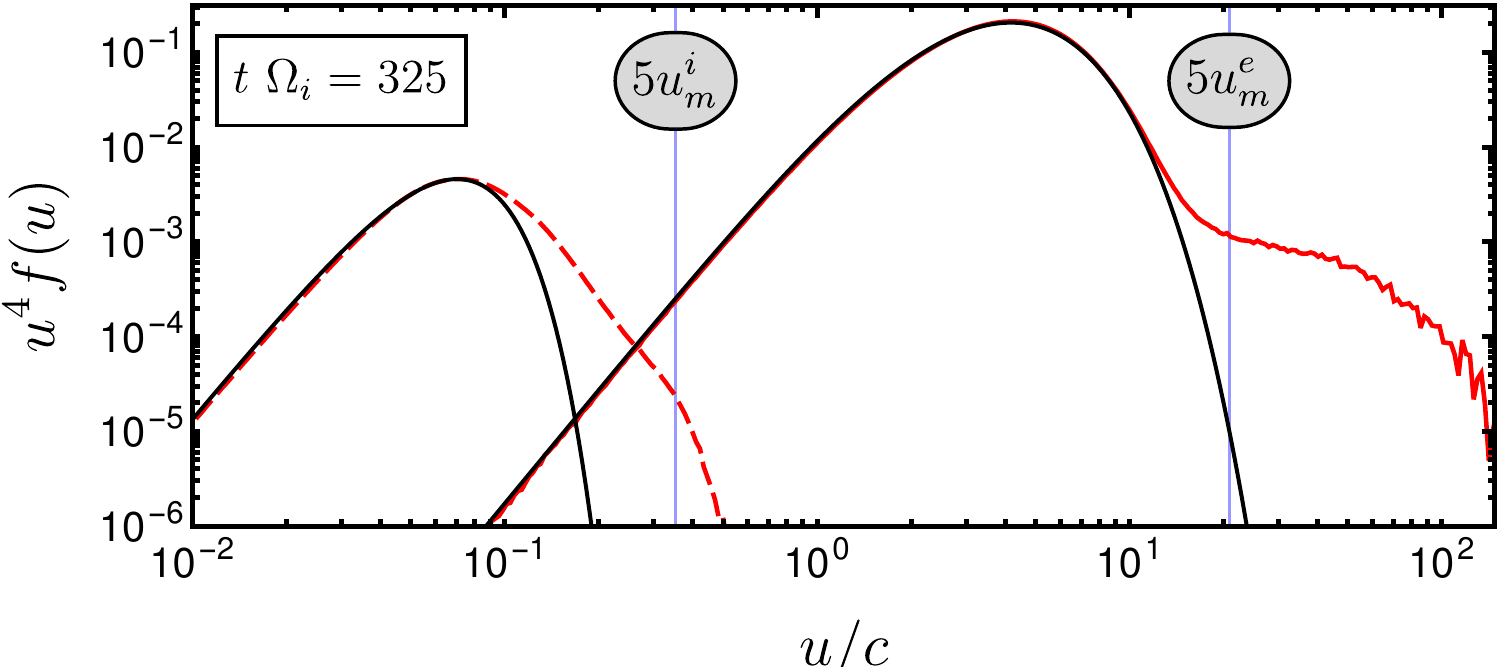}
\caption{
Momentum distribution for ions (red dashed) and electrons (red solid) at $t~ \Omega_i=325$ for the Ma5Mr1836 simulation, where we first transform momenta of particles to the plasma rest frame before we compute the momentum spectra.
Solid black lines show the analytical Maxwell-J{\"u}ttner distribution scaled appropriately, $u^4 f(u)$ (as given by Equation~\ref{Eq:distribution}). The normalized temperature (given by Equation~\ref{Eq:TR}) is determined from the location of the peaks ($u_m^i$ for ions and $u_m^e$ for electrons).
Particle energies with $u>5u_m$ (to the right of blue lines) are considered to be non-thermal.
\label{fig:Thermal}
}
\end{figure}

Here, we consider the full momentum distribution $f_{\rm full}$ that contains non-thermal and high-energy particles. 
The main assumption adopted here is that {\it{all low energy particles are in thermal equilibrium}}.
That is, all non-thermal particle are such that $u \gg u_m$, which means the value of $u$ at which the derivative in Equation~\eqref{Eq:du4fu} is zero is always outside the range for which the non-thermal parts of the distribution is non-zero.
Therefore, the value of $u_m$ and hence $\T_R$, computed from Equation~\eqref{Eq:TR}, is still the same.
An example of such a case is shown in Figure~\ref{fig:Thermal}.

Assuming that the non-thermal part of the distribution follows a power law with a typical index of -$4$, the full distribution can be approximated as follows,
\begin{eqnarray}
\label{Eq:Ffull}
f_{\rm full}(u) = \frac{n_R}{1+\alpha}
\left[
\frac{e^{-\gamma/\T_R}}{4 \pi k_2 (1/\T_R)}
+
\alpha
\frac{u_1 u_2 u^{-4}}{4 \pi (u_1 - u_2) }\theta_1 \theta_2
\right],
\end{eqnarray}
where $\alpha$ is the fractional number density of non-thermal particles which is implicitly assumed to be small, i.e., $\alpha \ll 1$, $\theta_1 = \theta(u-u_1)$, $\theta_2 = \theta(u_2 -u)$, and $u_1$ and $u_2$ ($u_2>u_1$) is the range in which the distribution takes a power-law form.
The distribution in Equation~\eqref{Eq:Ffull} is discontinuous at $u=u_1$. However, in reality the thermal and $u^{-4}$ parts of the distribution are connected with a suprathermal distribution (with a logarithmic slope steeper than $-4$) that starts at $u \sim 4 u_m$ to $ 5 u_m$ \citep{Caprioli2014a}.
Therefore, the suprathermal distribution does not change the relation between $u_m$ and $\T_R$ (Equation~\ref{Eq:TR}).

The inclusion of the non-thermal part leads to a lower fraction of particles in thermal equilibrium and hence lowers proportionately the value of $E_{\rm th}$ at a given $\T_R$. Namely
\begin{eqnarray}
\label{Eq:Eth2}
\langle E_{\rm th} \rangle_{\rm full}  \approx \frac{1}{1+\alpha} \left( 3 \T_R +\frac{K_1 \left(1/\T_R \right)}{K_2\left(1/\T_R \right)} - 1 \right) m_s c^2.
\end{eqnarray}

The value of $\alpha$ can be determined by comparing the value of $f(u_m)$ and $f_{\rm full}(u_m)$, where the value of 
$f(u_m)$ is computed from Equation~\eqref{Eq:distribution} at $u_m$ where the expression $u^4 f_{\rm full}(u)$ is maximized, and
$f_{\rm full}(u_m)$ is computed from the normalized histogram of the particles' normalized momenta ($u$) in the simulation.

\subsection{Acceleration efficiency}

We note here that $\alpha = n_{\rm non-th}/n_{\rm th}$, where $n_{\rm th}$ is the number density of particles in thermal equilibrium, and $n_{\rm non-th}$ is the number density of non-thermal particles. Thus, 
the rest-frame total number density is $n_R = n_{\rm th} + n_{\rm non-th}$.
That is, we can define the efficiency of acceleration as the fraction of non-thermal particles ($\epsilon_n$) measured in per cent via
\begin{eqnarray}
\label{Eq:EffN}
\epsilon_n \equiv \frac{n_{\rm non-th}}{n_R} \times 100 = \frac{\alpha}{1+\alpha} \times 100  .
\end{eqnarray}

We can also define the acceleration efficiency by the fractional energy carried by non-thermal particles ($\epsilon_E$) in per cent as
\begin{eqnarray}
\label{Eq:EffE}
\epsilon_E & \equiv & 
\frac{ E_{\rm tot} - \langle E_{\rm th} \rangle_{\rm full} }{ E_{\rm tot} } \times 100  
=
\frac{E_{\rm non-th}}{E_{\rm tot}} \times 100 ,
\end{eqnarray}
where $ E_{\rm tot} = \langle \gamma -1 \rangle m_s c^2$ is the average of $(\gamma-1)$ of all thermal and non-thermal particles.

An equally important definition of the acceleration efficiency is how much of the upstream plasma kinetic energy is channeled into non-thermal energy. In this case, the efficiency of acceleration in per cent is defined as
\begin{eqnarray}
\label{Eq:Eff-sh}
\epsilon_{\rm sh}
& \equiv &
\frac{ E_{\rm non-th}   }{ 
0.5  ~ (m_e + m_i) \varv_{\rm sh}^2 } \times 100 ,
\end{eqnarray}
where $\varv_{\rm sh}$ is the upstream plasma drifting speed in the shock rest-frame.
We compute the non-thermal energy density, $E_{\rm non-th}$, as the average (per particle) energy with $u>5u_m$ \citep[][]{Xu2020}. The vertical blue lines in Figure~\ref{fig:Thermal} show such cuts.
That is, particles with velocities $u>5u_m$ are assumed to form the non-thermal part of the distribution function.

\end{appendix}

\bibliography{refs}
\bibliographystyle{aasjournal}

\end{document}